\documentclass[aps,prd,twocolumn,nofootinbib]{revtex4-1}

\usepackage{graphicx}
\usepackage{hyperref}
\usepackage{slashed}
\usepackage{color}

\begin{document}

\title{Doubly Heavy Baryon Production at A High Luminosity $e^+ e^-$ Collider}
\author{Jun Jiang}
\author{Xing-Gang Wu}
\email{wuxg@cqu.edu.cn}
\author{Qi-Li Liao}
\author{Xu-Chang Zheng}
\author{Zhen-Yun Fang}

\address{Department of Physics, Chongqing University, Chongqing 401331, P.R. China}

\date{\today}

\begin{abstract}

Within the framework of nonrelativistic QCD, we make a detailed discussion on the doubly heavy baryon production through the $e^+ e^-$ annihilation channel, $e^{+}e^{-}\rightarrow\gamma/Z^0 \rightarrow \Xi_{QQ^{\prime}} +\bar{Q} +\bar{Q^{\prime}}$, at a high luminosity $e^{+}e^{-}$ collider. Here $Q^{(\prime)}$ stands for the heavy $b$ or $c$ quark. In addition to the channel through the usually considered diquark state $(QQ^{\prime})[^3S_1]_{\bf\bar{3}}$, contributions from the channels through other same important diquark states such as $(QQ^{\prime})[^1S_0]_{\bf 6}$ have also been discussed. Uncertainties for the total cross sections are predicted by taking $m_c=1.80\pm0.30$ GeV and $m_b=5.10\pm0.40$ GeV. At a super $Z$-factory running around the $Z^0$ mass and with a high luminosity up to ${\cal L} \propto 10^{34}\sim 10^{36}{\rm cm}^{-2} {\rm s}^{-1}$, we estimate that about $1.1\times10^{5 \sim 7}$ $\Xi_{cc}$ events,  $2.6\times10^{5 \sim 7}$ $\Xi_{bc}$ events and  $1.2\times 10^{4 \sim 6}$ $\Xi_{bb}$ events can be generated in one operation year. Such a $Z$-factory, thus, will provide a good platform for studying the doubly heavy baryons in comparable to the CERN large hadronic collider. \\

\noindent {\bf PACS numbers:} 13.66.Bc,12.38.Bx,14.20.-c

\end{abstract}

\maketitle

\section{Introduction}

Theoretically, the production of the doubly heavy baryons $\Xi_{QQ^{\prime}}$ has been analyzed in Refs.~\cite{Kiselev,chang,cqww,gencc1,gencc2,FLSW,K1,K2,K3,zhang,Michael,SPB,zhong,majp,sizongguo}, where the symbol $Q^{(\prime)}$ stands for the heavy $b$ or $c$ quark accordingly. In particular, a computer program GENXICC for simulating the hadronic production of the $\Xi_{cc}$, $\Xi_{bc}$ and $\Xi_{bb}$ has been completed and upgraded in Refs.~\cite{gencc1,gencc2}, which is written in a PYTHIA-compatible format~\cite{pythia}. Throughout the paper, $\Xi_{QQ'}$, corresponding to $\Xi_{cc}$, $\Xi_{bc}$ or $\Xi_{bb}$ respectively, is a short notation for the baryon $\Xi_{QQ'q}$ with the light quark $q$ equals to $u$ or $d$ or $s$ respectively \footnote{In the present paper, we will ignore the isospin-breaking effect, for instance, $\Xi_{cc}$ denotes $\Xi^+_{ccd}$ or $\Xi^{++}_{ccu}$ or $\Omega^+_{ccs}$ accordingly.}. Experimentally, among the doubly heavy baryons, only $\Xi_{cc}$ has been observed by SELEX collaboration~\cite{Mattson:2002vu,Moinester:2002uw,Ocherashvili:2004hi}. Neither Babar collaboration nor Belle collaboration have found the evidence for $\Xi_{cc}$ in related experiments~~\cite{Babar,Belle}, to say nothing of the $\Xi_{bc}$ and $\Xi_{bb}$ baryons. We hope the CERN large hadronic collider (LHC), due to its high collision energy and high luminosity, will change the present situation, and especially, improve our understanding on those baryons' hadronic production properties.

Comparing to the $pp$, $ep$ and $\gamma\gamma$ collisions, a $e^+e^-$ collider is helpful and has some advantages to perform precise measurements for certain processes. To seek the $\Xi_{QQ^{\prime}}$ events at the LHC is feasible, but its hadronic background is much noisy in comparison to a $e^+ e^-$ collider. In the hadronic production, the baryons are produced through scattering or annihilating or fusion of two initial partons inside the incident hadrons. In addition to the dominant gluon-gluon fusion mechanism, one also needs to take the extrinsic or intrinsic heavy quark mechanisms into consideration, especially at the small $p_t$ regions~\cite{cqww,chang}. Hence, the hadronic production becomes much more complicated due to the introduction of non-perturbative parton distribution functions and intrinsic components of the hadron (even though they are universal). While at the $e^{+}e^{-}$ collider, one only needs to consider the $e^{+}e^{-}$ annihilation channel, $e^{+}e^{-}\rightarrow\gamma/Z^0 \rightarrow \Xi_{QQ^{\prime}}+\bar{Q}+\bar{Q^{\prime}}$, which allows one to study $\Xi_{QQ^{\prime}}$ baryon's own properties.

If the luminosity of a $e^+e^-$ collider is ${\cal L}\propto 10^{34-36}cm^{-2}s^{-1}$ and its colliding energy is around the $Z^{0}$-peak (it is called as a super $Z$ factory~\cite{wjw}), it will raise the production rate up to several orders in comparison to the previous LEP and Belle and Babar experiments. This increment has already been observed in the doubly heavy meson production due to the $Z^0$-boson resonance effect~\cite{zbc0,zbc1,zbc11,zbc12,zbc2,zbc3,zbc4,zbc5,cc1}. It is thus natural to estimate that such a super $Z$-factory also opens new opportunities for studying the $\Xi_{QQ^{\prime}}$ baryon properties, such as their spectroscopy, their inclusive and exclusive decays, and etc.. In the present paper, we will study the semi-inclusive production of doubly heavy baryon $\Xi_{QQ^{\prime}}$ at the super $Z$-factory.

With the non-relativistic QCD (NRQCD) framework~\cite{nrqcd}, the production of $\Xi_{QQ^{\prime}}$ baryon can be factorized into two steps: The first step is to produce two free heavy-quark pairs $Q\bar{Q}$ and $Q^{\prime}\bar{Q}^{\prime}$, which is perturbative QCD (pQCD) calculable. This is due to the fact that the intermediate $\gamma$, gluon or $Z^0$ should be hard enough to generate a heavy-quark pair. The second step is to make the two heavy quarks $Q$ and $Q^{\prime}$ into a bounding diquark $(QQ^{\prime})$ in $[^3S_1]$ (or $[^1S_0]$) spin state and in $\mathbf{\bar{3}}$ (or $\mathbf{6}$) color state accordingly; then it will be hadronized into $\Xi_{QQ^{\prime}}$ baryon, whose probability is described by the NRQCD matrix element. More explicitly, the intermediate diquarks in $\Xi_{cc}$ and $\Xi_{bb}$ have two spin-and-color configurations $[^3S_1]_{\bf\bar{3}}$ and $[^1S_0]_{\bf 6}$; while for the intermediate diquark $(bc)$ in $\Xi_{bc}$, there are four spin-and-color configurations $\Xi_{bc}[^3S_1]_{\bf\bar{3}}$, $\Xi_{bc}[^3S_1]_{\bf 6}$, $\Xi_{bc}[^1S_0]_{\bf\bar{3}}$, and $\Xi_{bc}[^1S_0]_{\bf 6}$.

It has been observed that, for the hadronic production channels, the contributions from other spin and color configurations of the diquark such as $(QQ^{\prime})[^1S_0]_{\bf 6}$ configuration and etc. can also provide sizable contributions in addition to the dominant $(QQ^{\prime})[^3S_1]_{\bf \bar{3}}$~\cite{cqww,majp,sizongguo}. We will show that this is also the case for the present considered $e^{+}e^{-}$ annihilation channel, $e^{+}e^{-}\rightarrow\gamma/Z^0 \rightarrow \Xi_{QQ^{\prime}}+\bar{Q}+\bar{Q^{\prime}}$, thus, one needs to take all these states into consideration for a sound estimation.

The remaining parts of the paper are organized as follows. In Sec.II, we present the detailed formulation for dealing with the process of $e^{+}+e^{-}\rightarrow \gamma/Z^{0} \rightarrow \Xi_{QQ^{\prime}}+\bar{Q}+\bar{Q^{\prime}}$. In Sec.III, we give the numerical results and uncertainty discussion. Sec.IV is reserved for a summary.

\section{Calculation Technology}

\begin{figure}
\includegraphics[width=0.40\textwidth]{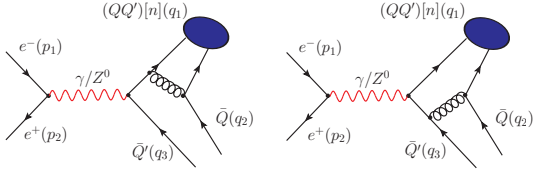}
\caption{Typical Feynman diagrams for $e^{+} e^{-}\rightarrow (QQ^{\prime})[n]+\bar Q+\bar Q^{\prime}$, where $Q$ and $Q^{\prime}$ stand for the heavy $c$ or $b$ quark, $[n]$ represents the spin-and-color quantum number of the intermediate diquark-state. Two new Feynman diagrams for the case of $Q\neq Q'$ can be obtained by exchanging quark-lines of $Q$ and $Q'$.} \label{feynman1}
\end{figure}

Typical Feynman diagrams for $e^{+}(p_2) e^{-}(p_1)\rightarrow (QQ^{\prime})[n](q_1)+\bar{Q}(q_2)+\bar Q^{\prime}(q_3)$ are presented in Fig.(\ref{feynman1}). According to NRQCD factorization formulae~\cite{petrelli}, its differential cross section can be written in the following factorization form:
\begin{eqnarray}
&d\sigma(e^{+} e^{-}\rightarrow \Xi_{QQ^{\prime}}+\bar{Q}+\bar{Q^{\prime}}) \nonumber\\
&=\sum\limits_{n} d\hat\sigma \left(e^{+} e^{-}\rightarrow (QQ^{\prime})[n]+\bar{Q}+\bar{Q^{\prime}}\right) \langle{\cal O}^H(n)\rangle, \label{total}
\end{eqnarray}
where the matrix element $\langle{\cal O}^H(n)\rangle$ is proportional to the inclusive transition probability of the perturbative state $(QQ^{\prime})[n]$ pair into the heavy baryon $\Xi_{QQ^{\prime}}$, $[n]$ represents the spin-and-color quantum numbers for the intermediate diquark state. The short-distance cross-section $d\hat\sigma(e^{+} e^{-}\rightarrow (QQ^{\prime})[n]+\bar{Q}+\bar{Q^{\prime}})$ takes the following form,
\begin{displaymath}
d\hat\sigma\left(e^{+} e^{-}\rightarrow (QQ^{\prime})[n]+\bar{Q}+\bar{Q^{\prime}}\right)  = \frac{\overline{\sum} |{\cal M}|^{2} d\Phi_3}{4\sqrt{(p_1\cdot p_2)^2-m_e^4}} , \label{cs}
\end{displaymath}
where ${\cal M}$ is the hard scattering amplitude, $\overline{\sum}$ means we need to average over the spin states of electron and positron and to sum over the color and spin of all final particles. The three-particle phase space is
\begin{displaymath}
d{\Phi_3}=(2\pi)^4 \delta^{4}\left(p_1+p_2 - \sum_f^3 q_{f}\right)\prod_{f=1}^3
\frac{d^3 {q_f}}{(2\pi)^3 2 q_f^0}.
\end{displaymath}
The phase-space can be generated and integrated with the help of the FormCalc program~\cite{formcalc}, or a combination of RAMBOS~\cite{rambos} and VEGAS~\cite{vegas} which can be found in generators GENXICC~\cite{gencc1,gencc2} and BCVEGPY~\cite{bcvegpy}. \\

\subsection{${\cal M}\left(e^{+} e^{-} \to (Q Q^{\prime})[n] + Q + \bar {Q^{\prime}}\right)$}

The hard scattering amplitude ${\cal M}$ for the diquark production can be related with the familiar meson production through a proper correspondence.

The hard scattering amplitude ${\cal M}$ for the production channel, $e^{+}(p_2) e^{-}(p_1)\rightarrow (QQ^{\prime})[n](q_1)+\bar{Q}(q_2)+\bar{Q^{\prime}}(q_3)$, can be written as :
\begin{widetext}
\begin{eqnarray}
{\cal M}\left((QQ^{\prime})[n]\right) &=& \bar{u}_{s_1}\left(\frac{m_Q}{M_{QQ^{\prime}}}q_1\right) \Gamma_{\rho}
s_f(k_{\rho-1},m_Q) \cdots s_f(k_1,m_Q) \Gamma_1 v_{s_4}(q_2) {\cal B}(S,s_1,s_2;q_{1},M_{QQ^{\prime}}) \nonumber\\
& &\times \bar{u}_{s_2}\left(\frac{m_{Q^{\prime}}}{M_{QQ^{\prime}}}q_1\right) \Gamma^\prime_1 s_f(k^\prime_1,m_{Q^{\prime}})  \cdots s_f(k^\prime_{\kappa-1},m_{Q^{\prime}})\Gamma^\prime_{\kappa} v_{s_3}(q_3) \times {\cal C} \times {\cal G} \times {\cal D} \times {\cal L}_{rr^{\prime}}  \;.   \label{diquarkline}
\end{eqnarray}
\end{widetext}
where $s_i$ is the spin state of the outgoing (anti)quark, $S$ is the spin state of the diquark, and $M_{QQ^{\prime}}$ is the diquark mass. The parameters $\Gamma_1$, $\cdots$, $\Gamma_\rho$, $\Gamma^\prime_1$, $\cdots$, $\Gamma^\prime_\kappa$ are sequential interaction vertexes (the $\gamma$-matrix elements only) along the corresponding spinor lines, $s_f(k^{(\prime)}_i,m_{Q^{(\prime)}})$ is the fermion propagator in between the interaction vertexes. As a special case, when there is only one interaction vertex ($\rho=1$ or $\kappa=1$), there is no fermion propagator. ${\cal B}(S,s_1,s_2;q_{1},M_{QQ^{\prime}})$ is the wavefunction of $(QQ^{\prime})[n]$-diquark. Here, ${\cal C}$ is the color factor of the process, ${\cal G}$ is the gluon propagator, ${\cal D}$ is the photon or $Z^{0}$ propagator, and ${\cal L}_{rr^{\prime}}$ is the leptonic part which can be expressed as
\begin{eqnarray*}
{\cal L}_{rr^{\prime}} = \bar{v}_r(p_2) \Gamma u_{r^{\prime}}(p_1) \;,
\end{eqnarray*}
in which $r^{(\prime)}$ stands for the spin state of the electron (positron), $\Gamma=\gamma^\mu$ for $\gamma$-propagator and $\Gamma=\gamma^\mu(\frac{1}{4}-\sin^2\theta_w-\frac{1}{4}\gamma^5)$
for $Z^{0}$-propagator.

As a comparison, the hard scattering amplitude for the meson $(Q^{\prime}\bar{Q})[n]$ production through the channel, $e^{+}(p_2) e^{-}(p_1)\to (Q^{\prime}\bar{Q})[n](q_1) + Q(q_2) + \bar {Q^{\prime}}(q_3)$, can be written as :
\begin{widetext}
\begin{eqnarray}
{\cal M}\left((Q^{\prime}\bar{Q})[n]\right) &=& \bar{u}_{s_4}(q_2)\Gamma_1 s_f(-k_1,m_Q) \cdots s_f(-k_{\rho-1}, m_Q) \Gamma_\rho v_{s_1}\left(\frac{m_Q}{M_{Q^{\prime}\bar{Q}}}q_1\right)
{\cal B}(S,s_1,s_2; q_1,M_{Q^{\prime}\bar{Q}}) \nonumber\\
&& \times\bar{u}_{s_2}\left(\frac{m_{Q^{\prime}}} {M_{Q^{\prime}\bar{Q}}}q_1\right) \Gamma^\prime_1 s_f(k^\prime_1,m_{Q^{\prime}}) \cdots s_f(k^\prime_{\kappa-1})\Gamma^\prime_{\kappa} v_{s_3}(q_3) \times {\cal C}^{\prime} \times {\cal G} \times {\cal D} \times {\cal L}_{rr^{\prime}} \;. \label{mesonline}
\end{eqnarray}
\end{widetext}
where $M_{Q^{\prime}\bar{Q}}$ is the meson mass, ${\cal B}(S,s_1,s_2;q_{1},M_{Q^{\prime}\bar{Q}})$ is the wavefunction of $(Q^{\prime}\bar{Q})[n]$-meson and ${\cal C}^{\prime}$ is the color factor for the present case. Other parameters have the same meaning as above.

A comparison of Eq.(\ref{diquarkline}) and Eq.(\ref{mesonline}) tells us that if we can transform the fermion line of the diquark-production case to an anti-fermion line of the meson-production case, then we can derive the diquark amplitude following the same way of meson production. More definitely, we need to deal with the following fermion line,
\begin{widetext}
\begin{displaymath}
a=\bar{u}_{s_1}\left(\frac{m_Q}{M_{QQ^{\prime}}}q_1\right) \Gamma_\rho s_f(k_{\rho-1},m_Q) \cdots s_f(k_1,m_Q) \Gamma_1 v_{s_4}(q_2) .
\end{displaymath}
\end{widetext}
Setting $T$ to be the operation of matrix transportation and $C=-i\gamma^2\gamma^0$ to be the charge conjugation matrix, we obtain the following equations
\begin{eqnarray*}
&& v^T_{s_4}(q_2)C =-\bar{u}_{s_4}(q_2),\; C^- \Gamma^T_i C = - \Gamma_i,\; CC^{-}=1,\\
&& C^- s^T_f(k_i,m_Q) C = s_f(-k_i,m_Q) ,\\
&& C^{-}\bar{u}^T_{s_1} \left(\frac{m_Q}{M_{QQ^{\prime}}}q_1\right) = v_{s_1}\left(\frac{m_Q}{M_{QQ^{\prime}}}q_1\right) .
\end{eqnarray*}
Then, the fermion line $a$ changes to
\begin{widetext}
\begin{eqnarray*}
a = a^T &=& v^T_{s_4}(q_2) \Gamma^T_1 s^T_f(k_1,m_Q) \cdots s^T_f(k_{\rho-1},m_Q) \Gamma^T_\rho \bar{u}^T_{s_1}\left(\frac{m_Q}{M_{QQ^{\prime}}}q_1\right) \\
&=& v^T_{s_4}(q_2) C C^- \Gamma^T_1 C C^- s^T_f(k_1,m_Q) C C^- \cdots
C C^- s^T_f(k_{\rho-1},m_Q) C C^-\Gamma^T_\rho CC^- \bar{u}^T_{s_1} \left(\frac{m_Q}{M_{QQ^{\prime}}} q_1\right) \\
&=&(-1)^{(\rho+1)} \bar{u}_{s_4}(q_2)\Gamma_1 s_f(-k_1,m_Q) \cdots s_f(-k_{\rho-1},m_Q) \Gamma_\rho v_{s_1}\left(\frac{m_Q}{M_{QQ^{\prime}}}q_1\right) .
\end{eqnarray*}
\end{widetext}

Thus, Eq.(\ref{diquarkline}) can be transformed as
\begin{widetext}
\begin{eqnarray}
{\cal M}((QQ^{\prime})[n]) &=& (-1)^{(\rho+1)}\bar{u}_{s_4}(q_2)\Gamma_1 s_f(-k_1,m_Q) \cdots s_f(-k_{\rho-1},m_Q) \Gamma_\rho v_{s_1}\left(\frac{m_Q} {M_{QQ^{\prime}}}q_1\right){\cal B}(S,s_1,s_2;q_1,M_{QQ^{\prime}})\nonumber \\
& & \qquad\qquad \times\bar{u}_{s_2}\left(\frac{m_{Q^{\prime}}} {M_{QQ^{\prime}}}q_1\right) \Gamma^\prime_1 s_f(k^\prime_1,m_{Q^{\prime}}) \cdots s_f(k^\prime_{\kappa-1}) \Gamma^\prime_{\kappa} v_{s_3}(q_3)\times {\cal C}\times {\cal G} \times {\cal D} \times {\cal L}_{rr^{\prime}}. \label{newline}
\end{eqnarray}
\end{widetext}
Comparing Eq.(\ref{mesonline}) with Eq.(\ref{newline}), one can, thus, follow the same procedures of meson production to finish the calculation.

By taking the Lorentz indexes and the color indexes explicitly, the hard scattering amplitude ${\cal M}$ in Eq.(\ref{newline}) can be rewritten as:
\begin{equation}
i{\cal M}((QQ^{\prime})[n]) = \kappa \sum\limits_{{\cal S} = 1}^4{ {\cal A}^{\mu}_{\cal S} } \times {\cal D}_{\mu\nu} \times {\cal L}^{\nu}_{rr^{\prime}}. \label{MM}
\end{equation}
For the production through $\gamma$-propagator: the overall parameter $\kappa$=$e_{Q^{(\prime)}} e^2 g_s^2 {\cal C}_{ij}$, where $e_{Q^{(\prime)}}$ is the electric charge, in unit of $e$, of the quark $Q^{(\prime)}$ and ${\cal C}_{ij}$ is the color factor with $i$ and $j$ the color indices of the outgoing antiquarks, the propagator ${\cal D}_{\mu\nu}=\frac{-i}{k^2}g_{\mu\nu}$ and the leptonic vector $${\cal L}_{rr^{\prime}}^{\nu}=\bar{v}_r(p_2) \gamma^\nu u_{r^{\prime}}(p_1).$$ For the production through $Z^{0}$-propagator: $\kappa=\frac{g^2 g_s^2}{\cos^2\theta_w} {\cal C}_{ij}$, $${\cal D}_{\mu\nu}=\frac{i}{k^2-m^2_Z +im_Z\Gamma_z}\left(-g_{\mu\nu}+\frac{k_\mu k_\nu}{k^2}\right)$$ with $\Gamma_z$ the total decay width of $Z^{0}$ boson, and $${\cal L}_{rr^{\prime}}^{\nu}=\bar{v}_r(p_2) \gamma^\nu\left( \frac{1}{4} - \sin^2\theta_w - \frac{1}{4}\gamma^5 \right) u_{r^{\prime}}(p_1).$$ The vectors ${\cal A}^{\mu}_{\cal S}$ (${\cal S}=1,\cdots,4$) can be read from the Feynman diagrams in Fig.(\ref{feynman1}). More explicitly, ${\cal A}^{\mu}_{\cal S}$ can be written as
\begin{widetext}
\begin{eqnarray}
{\cal A}^{\mu}_1 &=& {\bar u_s}({q_2}){\gamma_\rho}\frac{\Pi^{0(1)}_{(QQ^{\prime})[n]}(q_1)} {(q_{11} + {q_2})^2}{\gamma_\rho} \frac{\slashed{q}_1 +\slashed{q}_2 + m_{Q^{\prime}}}{(q_1 +q_2)^2 - m^2_{Q^{\prime}}}{\Gamma^{\mu}_{zQ^{\prime}}}v_{s^{\prime}}(q_3), \label{A1}
\end{eqnarray}
\begin{eqnarray}
{\cal A}^{\mu}_2 &=& {\bar u_s}({q_2}){\gamma_\rho}\frac{\Pi^{0(1)}_{(QQ^{\prime})[n]}(q_1)} {(q_{11} + {q_2})^2}{\Gamma^{\mu}_{zQ^{\prime}}} \frac{-\slashed{q_{11}}-\slashed{q}_2 -\slashed{q}_3 + {m_{Q^{\prime}}}}{(-\slashed{q_{11}}-\slashed{q}_2 -\slashed{q}_3)^2 - m^2_{Q^{\prime}}}{\gamma_\rho}v_{s^{\prime}}(q_3), \label{A2}
\end{eqnarray}
\begin{eqnarray}
{\cal A}^{\mu}_3 &=& -{\bar u_s}({q_2}){\Gamma^{\mu}_{zQ}}\frac{-\slashed{q}_1 -\slashed{q}_3 + m_{Q}}{(-q_1 -q_3)^2 - m^2_{Q}}{\gamma_\rho}\frac{\Pi^{0(1)}_{(QQ^{\prime})[n]}(q_1)} {(q_{12} + {q_3})^2}{\gamma_\rho} v_{s^{\prime}}(q_3)  \label{A3}
\end{eqnarray}
and
\begin{eqnarray}
{\cal A}^{\mu}_4 &=& -{\bar u_s}({q_2}){\gamma_\rho}\frac{\slashed{q_{12}}+\slashed{q}_3 +\slashed{q}_2 + {m_Q}}{(\slashed{q_{12}}+\slashed{q}_3 +\slashed{q}_2)^2 - m^2_{Q}}{\Gamma^{\mu}_{zQ}}\frac{\Pi^{0(1)}_{(QQ^{\prime})[n]}(q_1)} {(q_{12} + {q_3})^2} {\gamma_\rho}v_{s^{\prime}}(q_3) , \label{A4}
\end{eqnarray}
\end{widetext}
where for convenience, we introduce a general interaction vertex $\Gamma^{\mu}_{zQ^{(\prime)}}$, the needed ones are $$\Gamma^{\mu}_{zc} ={\gamma^\mu}\left[\alpha+\beta\left(\frac{1}{4} - \frac{2}{3}\sin^2\theta_w - \frac{1}{4}\gamma^5\right)\right]$$ or $$\Gamma^{\mu}_{zb} ={\gamma^\mu}\left[\alpha+\beta\left(\frac{1}{4} - \frac{1}{3}\sin^2\theta_w - \frac{1}{4}\gamma^5\right)\right].$$ Here $\alpha =1$ and $\beta=0$ for $\gamma-Q^{(\prime)}-Q^{(\prime)}$ vertex, and $\alpha =0$ and $\beta=1$ for $Z^{0}-Q^{(\prime)}-Q^{(\prime)}$-vertex, respectively. The momenta of the constituent quarks which forms the bound state are
\begin{equation}
q_{11} = \frac{m_Q}{M_{QQ^{\prime}}}{q_1} + q \;\;{\rm and}\;\;
q_{12} = \frac{m_{Q^{\prime}}}{M_{QQ^{\prime}}}{q_1} - q,
\end{equation}
where $M_{QQ^{\prime}}= m_Q + m_{Q^{\prime}}$ is implicitly adopted to ensure the gauge invariance of the hard scattering amplitude, and $q$ is the relative momentum between the two constituent quarks inside the diquark. Due to the non-relativistic approximation, $q$ is small and neglected in the amplitude \footnote{The integration over $q$ results in a wavefunction at zero which has been absorbed into the overall non-perturbative matrix element.}. For the production of $(cc)$- and $(bb)$-diquarks, we only need to calculate $A^{\mu}_1$ and $A^{\mu}_2$; but need to time the squared amplitude for $(cc)$- and $(bb)$-diquark by an overall factor $(2^2/2!)=2$, where the $1/2!$ factor is due to the symmetry of the diquark wavefunction. While for the case of $(bc)$-diquark production, we need to calculate $A^{\mu}_{\cal S}$ (${\cal S}=1,\cdots,4$).

The projector $\Pi^{0(1)}_{(QQ^{\prime})[n]}(q_1)$, under the non-relativistic approximation, takes the following form~\cite{petrelli} :
\begin{equation}\label{eq:projector}
\Pi^{0(1)}_{(QQ^{\prime})[n]}(q_1)=\frac{1}{2\sqrt{M_{QQ^{\prime}}}} \left(\xi_1\gamma^{5}+\xi_2\slashed{\epsilon}(q_1)\right) (\slashed{q}_1+M_{QQ^{\prime}}) ,
\end{equation}
where $\epsilon(q_1)$ is the polarization vector for spin-triplet state. Here the projector $\Pi^{1}$ is for the case of spin-triplet $[^3S_1]$, which corresponds to $\xi_1 =0$ and $\xi_2=1$, and the projector $\Pi^{0}$ is for the case of spin-singlet $[^1S_0]$, which corresponds to $\xi_1 =1$ and $\xi_2=0$.

\subsection{The color factor $C_{ij}$}

Because of the fact that $3 \bigotimes 3={\bf \bar{3}} \bigoplus {\bf 6}$ in $SU_{C}(3)$ color group, the $(QQ')$-diquark can be either in anti-triplet ${\bf \bar{3}}$ or in sextuplet ${\bf 6}$ color state. According to Fig.(\ref{feynman1}), the color factor $C_{ij}$ of the process is
\begin{eqnarray}
{\cal C}_{ij}={\cal N}\times\sum_{m,n}(T^{a})_{im}(T^{a})_{jn} \times G_{mnk}
\end{eqnarray}
where $i,j,m,n=1,2,3$ are color indices of the two outgoing anti-quarks
$\bar{Q}$ and $\bar{Q}^{\prime}$ and the two constituent quarks ${Q}$ and ${Q}^{\prime}$ of the diquark respectively, $k$ is the color indices of the diquark $(QQ^{\prime})$, and the indices $a=1,\cdots,8$ is the color index for the gluon, and the normalization constant ${\cal{N}}=\sqrt{1/2}$. The function $G_{mnk}$ equals to the anti-symmetric function $\varepsilon_{mjk}$ and equals to the symmetric function $f_{mjk}$ accordingly: the anti-symmetric $\varepsilon_{mjk}$ satisfies
\begin{displaymath}
\varepsilon_{mjk}\varepsilon_{m'j'k}=\delta_{mm'}\delta_{jj'}-\delta_{mj'}\delta_{jm'} ,
\end{displaymath}
and the symmetric $f_{mjk}$ satisfies
\begin{displaymath}
f_{mjk}f_{m'j'k}=\delta_{mm'}\delta_{jj'}+
\delta_{mj'}\delta_{jm'}.
\end{displaymath}
Then, the square of ${\cal C}^2_{ij}$ equals to $\frac{4}{3}$ for the color anti-triplet diquark production, and $\frac{2}{3}$ for the color sextuplet diquark production, respectively.

\subsection{Transition from $(QQ^{\prime})[n]$ to $\Xi_{QQ^{\prime}}$}

According to NRQCD, the $\Xi_{QQ^{\prime}}$ can be expanded as a series of Fock states which are according to the relative velocity ($v$) of the constituent heavy quarks in the baryon rest frame :
\begin{eqnarray*}
\vert\Xi_{QQ^{\prime}} \rangle = c_1 \vert (QQ^{\prime}) q\rangle +c_2 \vert (QQ^{\prime}) qg \rangle +c_3 \vert (QQ^{\prime}) q gg \rangle +\cdots,
\end{eqnarray*}
where $c_i(i=1,2,\cdots)$ is a function of the small velocity $v$, where $v$ is the relative velocity of the heavy quarks in the rest frame of the diquark. As for $\Xi_{cc}$ and $\Xi_{bb}$, the intermediate binding diquark in the Fock states can be in either $(QQ)[^3S_1]_{\bf\bar{3}}$ or $(QQ)[^1S_0]_{\bf 6}$ ($Q=c, b$) state; while for $\Xi_{bc}$, the intermediate binding diquark within the Fock states can be in $(bc)[^3S_1]_{\bf\bar{3}}$ or $(bc)[^1S_0]_{\bf\bar{3}}$ or $(bc)[^3S_1]_{\bf 6}$ or $(bc)[^1S_0]_{\bf 6}$ states respectively. Ref.\cite{majp} has suggested that the baryon can be formed with the component $\vert (QQ^{\prime}) qg \rangle$ as well as the usual $\vert (QQ^{\prime}) q \rangle$: One of the heavy quarks emits a gluon, which does not change the spin of the heavy quark, and this gluon splits into a $q\bar q$, the light quarks can also emit gluons, then the component can be formed with the light quark $q$ plus one gluon. Because a light quark can emit gluons easily, one can take the transition probability from these diquark states into the corresponding baryon to be at the same importance. Then, as a rough order estimation, one can take the transition probability from these diquark states into the corresponding baryon to be the same; i.e.
\begin{equation}\label{approxh}
h_{\bf 6} \simeq h_{\bf\bar{3}} ,
\end{equation}
where $h_{\bf\bar{3}}$ stands for the probability of transforming the color anti-triplet diquark into the baryon and $h_{\bf 6}$ stands for the probability of transforming the color sextuplet diquark into the baryon. These non-perturbative matrix elements can be determined from the potential model or from the non-perturbative methods such as QCD sum rules and lattice QCD.

It is also noted in Ref.~\cite{majp} that if the baryon is formed by the Fock state component $|(QQ^{\prime})q\rangle$ only, the emitted gluon for the case of $(QQ^{\prime})$ in $^1S_0$ spin-state, which will split into a $q\bar{q}$-pair with $q$ being combined by $(QQ^{\prime})$ to form the baryon, must change the spin of the heavy quark. Then, according to NRQCD~\cite{nrqcd}, one may conclude that $h_{\bf 6}$ must be $v^2$-suppressed in comparison with $h_{\bf\bar{3}}$. This provides the underlying reason why only $h_{\bf \bar{3}}$ has been taken into consideration for the doubly heavy baryon production, c.f. Refs.~\cite{FLSW,K1,K2,Michael,SPB}. Fortunately, these matrix elements are overall parameters and their uncertainties can be conveniently discussed. We will adopt the approximation (\ref{approxh}) throughout the paper \footnote{In different to the heavy quark fragmentation, during the fragmentation of diquark into baryon, the diquark may dissociate, which will decrease the baryon production cross section to a certain degree. In our present calculation we will not take this effect into consideration. Thus, our present estimations can be treated as an upper limit for the total cross sections. }.

Furthermore, the color-singlet nonperturbative matrix element $\langle{\cal O}^{H} (1S) \rangle$ in the factorization formulae (\ref{total}) can be related to the Schr\"{o}dinger wavefunctions at the origin $|\psi_{(Q\bar{Q'})}(0)|$~\cite{petrelli}
\begin{eqnarray}
h_{\bf 6} \simeq h_{\bf\bar{3}}=\langle{\cal O}^{H} (1 S) \rangle \simeq |\psi_{|(Q\bar{Q'})[1 S]\rangle}(0)|^2.
\end{eqnarray}
Since the spin-splitting effect is small, we do not distinguish the bound-state parameters for the spin-singlet and the spin-triplet states; i.e. those parameters, such as the constituent quark masses, the bound state mass, the wavefunction and etc, are taken to be the same for the spin-singlet and the spin-triplet states.

\section{Numerical results}

When doing the numerical calculation, the input parameters are taken as the following values~\cite{cqww}:
\begin{eqnarray*}
& & |\Psi_{cc}(0)|^2=0.039 \; {\rm GeV}^3,\; \\
& & |\Psi_{bc}(0)|^2=0.065 \; {\rm GeV}^3,\; \\
& & |\Psi_{bb}(0)|^2=0.152 \; {\rm GeV}^3 ,  \\
& & m_c=1.8 \; {\rm GeV},\; m_b=5.1 \; {\rm GeV} ,\; \\
& & M_{\Xi_{cc}}=3.6 \; {\rm GeV},\; M_{\Xi_{bc}}=6.9 \; {\rm GeV},\; M_{\Xi_{bb}}=10.2 \; {\rm GeV} \ .
\end{eqnarray*}
Other input parameters are taken from the Particle Data Group~\cite{pdg}: $\Gamma_z=2.4952$ GeV, $m_Z=91.1876$ GeV, $m_W=80.399$ GeV, and ${\cos}{\theta _w}={m_W}/{m_Z}$. We take the renormalization scale to be $2m_c$ for $\Xi_{cc}$ or $\Xi_{bc}$, and $2m_b$ for $\Xi_{bb}$, which leads to $\alpha_s(2m_c)=0.212$ and $\alpha_s(2m_b)=0.164$ for leading-order $\alpha_s$ running.

\begin{table}[tb]
\begin{tabular}{|c||c|}
\hline
~~~ $$ ~~~ & ~~cross section (pb)~~ \\
\hline\hline
$e^+ e^-\to \gamma \to \Xi_{cc}([^3S_1]_{\bf\bar{3}})$  &  8.90 $\times$ $10^{-4}$\\
\hline
$e^+ e^-\to \gamma \to \Xi_{cc}([^1S_0]_{\bf 6}) $ &  4.29 $\times$ $10^{-4}$\\
\hline
$e^+ e^-\to Z^{0} \to \Xi_{cc}([^3S_1]_{\bf\bar{3}}) $  &  0.727\\
\hline
$e^+ e^-\to Z^{0} \to \Xi_{cc}([^1S_0]_{\bf 6}) $ &  0.353\\
\hline\hline
$e^+ e^-\to \gamma \to \Xi_{bc}([^3S_1]_{\bf\bar{3}}) $  & 2.66 $\times$ $10^{-4}$\\
\hline
$e^+ e^-\to \gamma \to \Xi_{bc}([^3S_1]_{\bf 6})$  & 1.33 $\times$ $10^{-4}$\\
\hline
$e^+ e^-\to \gamma \to \Xi_{bc}([^1S_0]_{\bf \bar{3}})$  &  1.85 $\times$ $10^{-4}$\\
\hline
$e^+ e^-\to \gamma \to \Xi_{bc}([^1S_0]_{\bf 6})$  &  9.27 $\times$ $10^{-5}$\\
\hline
$e^+ e^-\to Z^{0} \to \Xi_{bc}([^3S_1]_{\bf\bar{3}})$ & 1.00\\
\hline
$e^+ e^-\to Z^{0} \to \Xi_{bc}([^3S_1]_{\bf 6})$ & 0.502\\
\hline
$e^+ e^-\to Z^{0} \to \Xi_{bc}([^1S_0]_{\bf \bar{3}})$  & 0.730\\
\hline
$e^+ e^-\to Z^{0} \to \Xi_{bc}([^1S_0]_{\bf 6})$  & 0.365\\
\hline\hline
$e^+ e^-\to \gamma \to \Xi_{bb}([^3S_1]_{\bf\bar{3}})$  &  1.94 $\times$ $10^{-5}$\\
\hline
$e^+ e^-\to \gamma \to \Xi_{bb}([^1S_0]_{\bf 6})$  &  8.95 $\times$ $10^{-6}$\\
\hline
$e^+ e^-\to Z^{0} \to \Xi_{bb}([^3S_1]_{\bf\bar{3}})$  & 8.03 $\times$ $10^{-2}$\\
\hline
$e^+ e^-\to Z^{0} \to \Xi_{bb}([^1S_0]_{\bf 6})$  &  3.86 $\times$ $10^{-2}$\\
\hline
\end{tabular}
\caption{Total cross section (in unit: pb) for the production of $\Xi_{QQ^{\prime}}(n)$ baryons ($n$ stands for the intermediate diquark state) through $e^+ e^-$ annihilation at the $Z^{0}$-Peak ($\sqrt{S}=m_Z$). }
\label{crosssection}
\end{table}

At the super $Z$-factory, total cross sections for the production channel, $e^{+}+e^{-}\rightarrow \Xi_{QQ^{\prime}}(n)+\bar{Q}+\bar{Q^{\prime}}$ ($n$ stands for the intermediate diquark state), are presented in Table~\ref{crosssection}. It is found that,

\begin{itemize}
\item When the $e^+e^-$ collision energy is at the $Z^{0}$-peak ($E_{cm}=\sqrt{S}=m_Z$), the production channel through a $\gamma$-propagator is much smaller than the case of a $Z^0$-propagator; i.e. for the production through the same diquark state, its cross-section is always less than $10^{-3}$ of the $Z^0$-case.

\item At the $Z^{0}$-peak, the contribution from the diquark state $n=[^1S_0]_{\bf 6}$ configuration is almost half of that of $n=[^3S_1]_{\bf\bar{3}}$ for the $\Xi_{cc}$ and $\Xi_{bb}$ baryon production. While for the $\Xi_{bc}$ baryon production, four intermediate $(bc)$-diquark states will make sizable contributions: the total cross-sections of $n=[^3S_1]_{\bf 6}$, $[^1S_0]_{\bf \bar{3}}$ and $[^1S_0]_{\bf 6}$ are about $50\%$, $73\%$ and $37\%$ of that of $n=[^3S_1]_{\bf \bar{3}}$. Then, one should take all these Fock states' contributions into consideration in order to derive a sound estimation.
\end{itemize}

\begin{figure}
\includegraphics[width=0.40\textwidth]{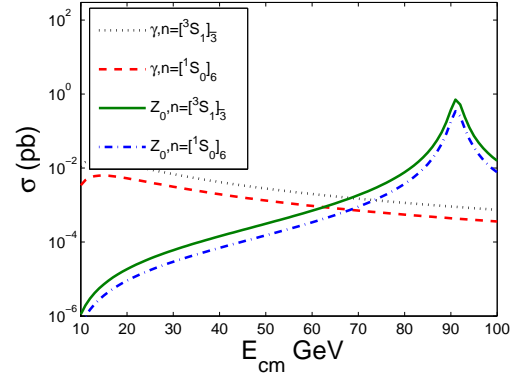}
\caption{Total cross sections (in unit: pb) of the production channel, $e^{+}+e^{-}\rightarrow \gamma/Z^{0} \rightarrow \Xi_{cc}(n)+\bar{c}+\bar{c}$, versus the $e^+ e^-$ collision energy $E_{cm}=\sqrt{S}$, where $n$ stands for the corresponding intermediate $(cc)$-diquark state. } \label{cc}
\end{figure}

\begin{figure}
\includegraphics[width=0.40\textwidth]{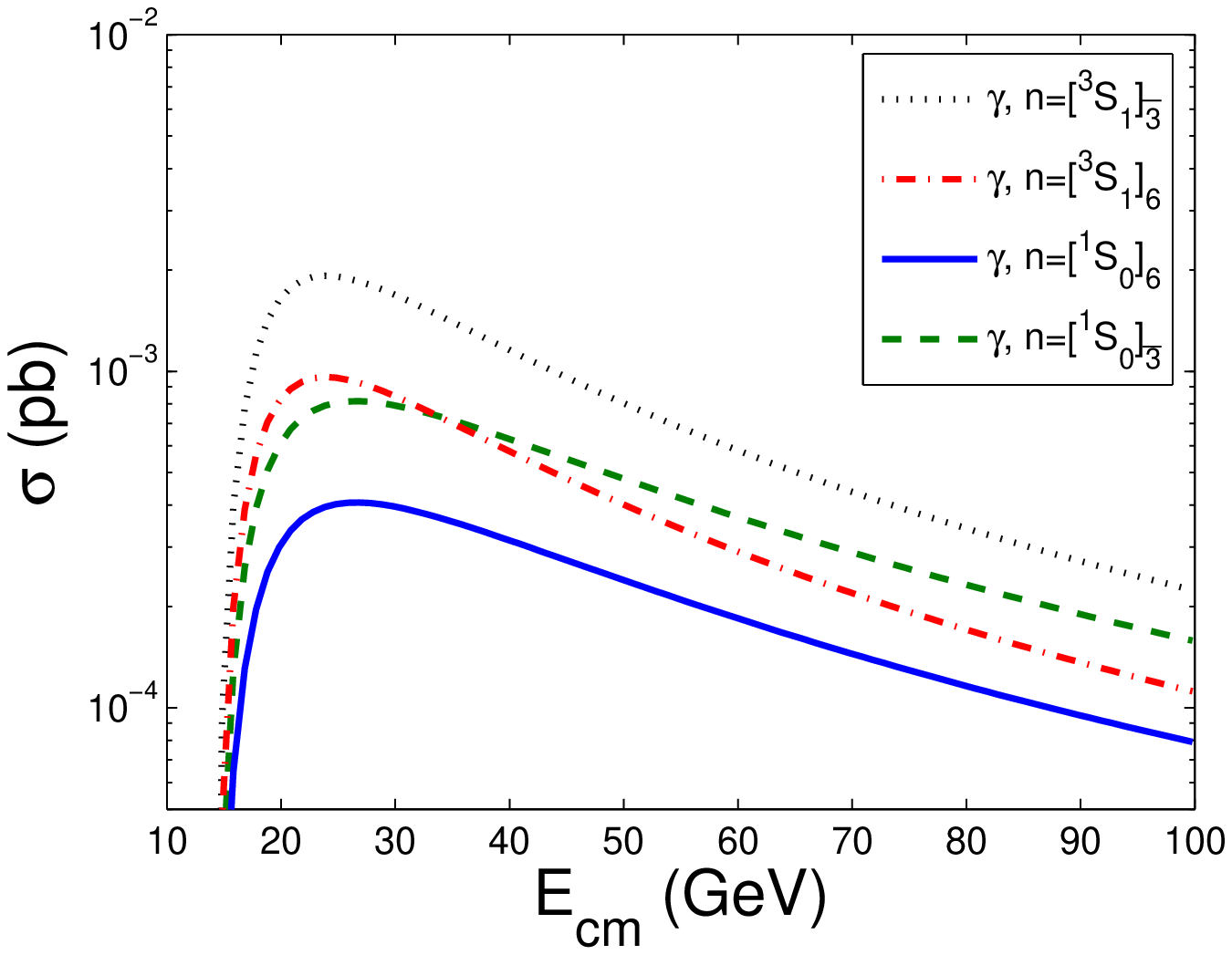}
\includegraphics[width=0.40\textwidth]{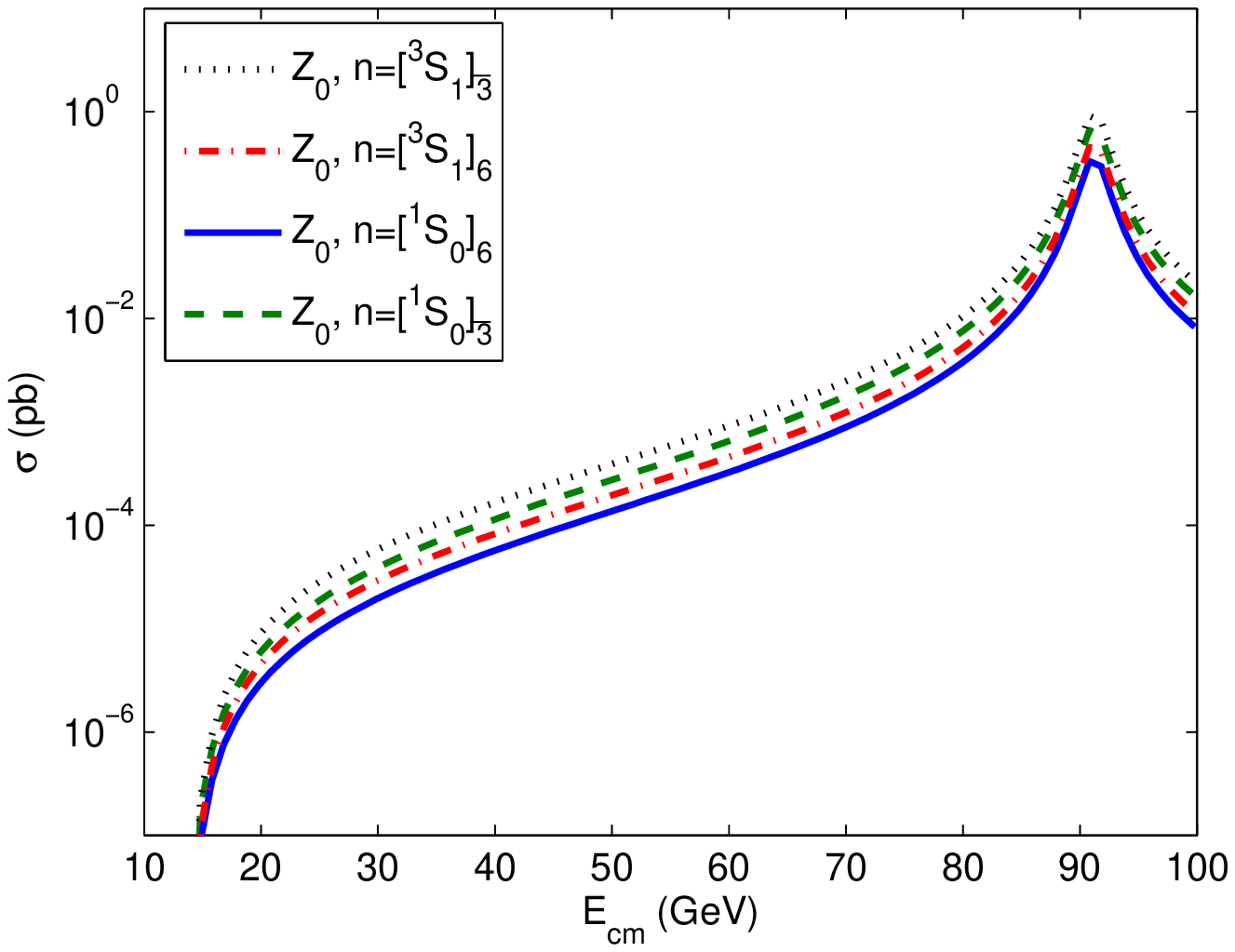}
\caption{Total cross sections (in unit: pb) of the production channel, $e^{+}+e^{-}\rightarrow \gamma/Z^{0} \rightarrow \Xi_{bc}(n)+\bar{c}+\bar{b}$, versus the $e^+ e^-$ collision energy $E_{cm}=\sqrt{S}$, where $n$ stands for the corresponding intermediate $(bc)$-diquark state. } \label{bc}
\end{figure}

\begin{figure}
\includegraphics[width=0.40\textwidth]{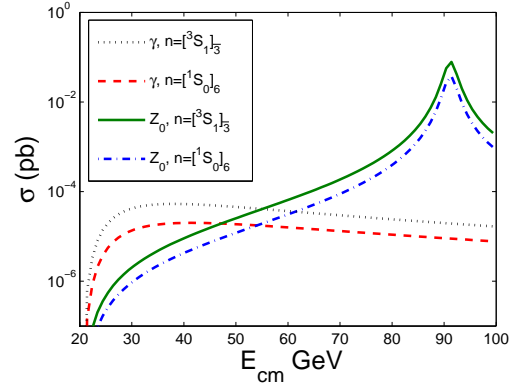}
\caption{Total cross sections (in unit: pb) of the production channel, $e^{+}+e^{-}\rightarrow \gamma/Z^{0} \rightarrow \Xi_{bb}(n)+\bar{b}+\bar{b}$, versus the $e^+ e^-$ collision energy $E_{cm}=\sqrt{S}$, where $n$ stands for the corresponding intermediate $(bb)$-diquark state.  } \label{bb}
\end{figure}

To show how the total cross sections change with the variation of $e^+e^-$ collision energy $E_{cm}=\sqrt{S}$, we present their cross sections versus the collision energy in Figs.(\ref{cc},\ref{bc},\ref{bb}) for the $\Xi_{cc}$, $\Xi_{bc}$ and $\Xi_{bb}$ baryon production respectively. In small collision energy region (e.g. $E_{cm}\lesssim 50 $ GeV), we have $$\sigma_{e^{+}e^{-}\to\gamma\to\Xi_{QQ^{\prime}}(n)} > \sigma_{e^{+}e^{-}\to Z^{0}\to \Xi_{QQ^{\prime}}(n)}.$$ The production cross section for the case of $\gamma$-propagator dominates at small $E_{cm}$, there is a small peak at $10 \sim 30$ GeV, and then, it decreases slowly with the increment of $E_{cm}$. The production cross section for the case of $Z^0$-propagator is negligible at small $E_{cm}$, but it arises logarithmically, and when $E_{cm}\sim m_{Z}$, due to the $Z^0$-resonance effect, it is about four-orders bigger than its value at $E_{cm}=50 $ GeV and is about three-orders bigger than the total cross-section (summed up cross-section for both the $\gamma$-propagator and the $Z^0$-propagator) in low region of $E_{cm}\sim 10-30$ GeV. This shows that a $e^+e^-$ collider running around the $Z^0$-peak will provide us a better chance for producing more doubly heavy baryons than the previous LEP, Babar and Belle platforms.

\begin{table}
\begin{tabular}{|c||c|c|c|c|c|}
\hline
  ~~~$E_{cm}$~~~    & ~$95\% m_Z$~& ~$97\% m_Z$~   & ~$m_Z$~   & ~$103\% m_Z$~ & ~$105\% m_Z$~ \\
\hline\hline
$\sigma_{\Xi_{cc}([^3S_1]_{\bf\bar{3}})}$ & 0.0487 & 0.1215 & 0.7283 & 0.1307 & 0.0543 \\
\hline
$\sigma_{\Xi_{cc}([^1S_0]_{\bf 6})}$ & 0.0236 & 0.0589 & 0.3533 & 0.0634 & 0.0263 \\
\hline\hline
$\sigma_{\Xi_{bc}([^3S_1]_{\bf\bar{3}})}$ & 0.0658 & 0.166 & 1.00 & 0.180 & 0.0744 \\
\hline
$\sigma_{\Xi_{bc}([^3S_1]_{\bf 6})}$ & 0.0329 & 0.0830 & 0.502 & 0.0899 & 0.0372 \\
\hline
$\sigma_{\Xi_{bc}([^1S_0]_{\bf \bar{3}})}$ & 0.0478 & 0.121 & 0.730 & 0.131 & 0.0542 \\
\hline
$\sigma_{\Xi_{bc}([^1S_0]_{\bf 6})}$ & 0.0239 & 0.0603 & 0.365 & 0.0655 & 0.0271 \\
\hline\hline
$\sigma_{\Xi_{bb}([^3S_1]_{\bf\bar{3}})}$ & 0.0052 & 0.0132 & 0.0803 & 0.0145 & 0.0060 \\
\hline
$\sigma_{\Xi_{bb}([^1S_0]_{\bf 6})}$ & 0.0025 & 0.0063 & 0.0386 & 0.0070 & 0.0029 \\
\hline
\end{tabular}
\caption{The cross section (in unit: pb) for the $\Xi_{QQ^{\prime}}$ baryon through the process $e^{+}e^{-} \rightarrow\gamma/Z^{0}\rightarrow \Xi_{QQ^{\prime}}(n)+\bar{Q}+\bar{Q^{\prime}}$ ($n$ stands for the intermediate diquark state) under several typical collision energies. } \label{diffmZ}
\end{table}

To be useful reference, we calculate the total cross section (summed up cross-section for both the $\gamma$-propagator and the $Z^0$-propagator) by taking the collision energy $E_{cm}=(1 \pm 3\%) m_Z$ and $E_{cm}=(1 \pm 5\%) m_Z$, which are shown by Table~\ref{diffmZ}. It shows that when the collision energy $E_{cm}$ is away from its center value of $m_Z$, the total cross sections drops down quickly: a $5\%$ deviation will lead to about one-order-lower cross section. Taking the production of ${\Xi_{cc}[^3S_1]_{\bf\bar{3}}}$ as an example, the total cross section for $E_{cm}=(1 \pm 3\%) m_Z$ will be lowered to $\sim\left({}^{18.0\%}_{16.7\%}\right)$ of its peak value; and the total cross section for $E_{cm}=(1 \pm 5\%) m_Z$ will be lowered to $\sim\left({}^{7.5\%}_{6.7\%}\right)$ of its peak value.

Considering the super $Z$-factory which is running around the mass of $Z^{0}$-boson and with a high luminosity ${\cal L}\propto 10^{34} {\rm cm}^{-2}{\rm s}^{-1}$, the number of $\Xi_{QQ^{\prime}}$($Q,Q^{\prime}=c,b$) events per year can be estimated \footnote{Approximately, 1 year $\approx \pi\times 10^7$ s, but it is common that a collider only operates about $1/\pi$ of the time a year~\cite{hantao}, i.e. we are customary to take $10^{34}\ {\rm cm}^{-2}\ {\rm s}^{-1}\approx 10^{5}{\rm pb}^{-1}\ {\rm /year}.$} :

\begin{itemize}
\item Totally, $1.08 \times 10^5$ $\Xi_{cc}$-baryon events can be generated, which includes $7.28 \times 10^4$ events coming from the intermediate $(cc)[^3S_1]_{\bf\bar{3}}$ diquark state and $3.53 \times 10^4$ events coming from the intermediate $(cc)[^1S_0]_{\bf 6}$ diquark state.

\item Totally, $2.60 \times 10^5$ $\Xi_{bc}$-baryon events can be generated, which includes $1.00 \times 10^5$ events coming from the intermediate $(bc)[^3S_1]_{\bf\bar{3}}$ diquark state, $5.02 \times 10^4$ events coming from the intermediate $(bc)[^3S_1]_{\bf 6}$ diquark state, $7.30 \times 10^4$ events coming from the intermediate $(bc)[^1S_0]_{\bf\bar{3}}$ diquark state and $3.65 \times 10^4$ events coming from the intermediate $(bc)[^1S_0]_{\bf 6}$ diquark state.

\item Totally, $1.19 \times 10^4$ $\Xi_{bb}$-baryon events can be generated, which includes $8.03 \times 10^3$ events coming from the intermediate $(bb)[^3S_1]_{\bf\bar{3}}$ diquark state and $3.86 \times 10^3$ events coming from the intermediate $(bb)[^1S_0]_{\bf 6}$ diquark state.

\item About $10^{4}-10^{5}$-order events per year shows clearly that sizable events of $\Xi_{cc}$, $\Xi_{bc}$, and even $\Xi_{bb}$ can be produced at the future super $Z$-factory. If its luminosity can be increased up to ${\cal L}\propto 10^{36} {\rm cm}^{-2}{\rm s}^{-1}$, the event numbers will be further increased by two orders. Thus, even if we have a slight deviation of $E_{cm}$ from $m_Z$, e.g. up to $5\%$-departure, there is still sizable events.
\end{itemize}

\begin{table}
\begin{tabular}{|c||c|c|c|c|c|}
\hline
~~$m_c$ ({\rm GeV})~~        & ~~1.50~~   & ~~1.65~~   & ~~1.80~~   & ~~1.95~~   & ~~2.10~~  \\
\hline \hline
$\sigma_{\Xi_{cc}([^3S_1]_{\bf\bar{3}})}$ & 1.270 & 0.949 & 0.727 & 0.569 & 0.453 \\
\hline
$\sigma_{\Xi_{cc}([^1S_0]_{\bf 6})}$ & 0.616 & 0.461 & 0.353 & 0.276 & 0.220 \\
\hline\hline
$\sigma_{\Xi_{bc}([^3S_1]_{\bf\bar{3}})}$ & 1.85 & 1.34 & 1.00 & 0.767 & 0.598 \\
\hline
$\sigma_{\Xi_{bc}([^3S_1]_{\bf 6})}$ & 0.926 & 0.672 & 0.502 & 0.383 & 0.299 \\
\hline
$\sigma_{\Xi_{bc}([^1S_0]_{\bf\bar{3}})}$ & 1.28 & 0.954 & 0.730 & 0.571 & 0.455 \\
\hline
$\sigma_{\Xi_{bc}([^1S_0]_{\bf 6})}$ & 0.639 & 0.477 & 0.365 & 0.285 & 0.227 \\
\hline
\end{tabular}
\caption{Uncertainties for the total cross section (in unit: pb) of the production channel $e^{+}e^{-} \rightarrow Z^{0} \rightarrow \Xi_{QQ^{\prime}}(n)+\bar{Q}+\bar{Q^{\prime}}$ ($n$ stands for the intermediate diquark state) with varying $m_c$. Here, $m_b$ is fixed to be $5.10$ GeV. }\label{uncertaintyz0c}
\label{tabmc}
\end{table}

\begin{table}
\begin{tabular}{|c||c|c|c|c|c|}
\hline ~~$m_b$ ({\rm GeV})~~    & ~~4.70~~   & ~~4.90~~   & ~~5.10~~   & ~~5.30~~   & ~~5.50~~  \\
\hline \hline
$\sigma_{ \Xi_{bc}([^3S_1]_{\bf\bar{3}})}$ & 0.987 & 0.995 & 1.00 & 1.01 & 1.02 \\
\hline
$\sigma_{ \Xi_{bc}([^3S_1]_{\bf 6})}$ & 0.494 & 0.498 & 0.502 & 0.506 & 0.509 \\
\hline
$\sigma_{\Xi_{bc}([^1S_0]_{\bf \bar{3}})}$ & 0.738 & 0.734 & 0.730 & 0.726 & 0.723 \\
\hline
$\sigma_{\Xi_{bc}([^1S_0]_{\bf 6})}$ & 0.369 & 0.367 & 0.365 & 0.363 & 0.362 \\
\hline\hline
$\sigma_{\Xi_{bb}([^3S_1]_{\bf\bar{3}})}$ & 0.105 & 0.092 & 0.080 & 0.071 & 0.062 \\
\hline
$\sigma_{\Xi_{bb}([^1S_0]_{\bf 6})}$ & 0.051 & 0.044 & 0.039 & 0.034 & 0.030 \\
\hline
\end{tabular}
\caption{Uncertainties for the total cross section (in unit: pb) of the production channel $e^{+}e^{-}\rightarrow Z^{0} \rightarrow \Xi_{QQ^{\prime}}(n)+\bar{Q}+\bar{Q^{\prime}}$ ($n$ stands for the intermediate diquark state) with varying $m_b$. Here, $m_c$ is fixed to be $1.80$ GeV. }\label{uncertaintyz0b}
\label{tabmb}
\end{table}

Finally, we discuss the theoretical uncertainties from the heavy quark masses by varying $m_c=1.80\pm0.30$ GeV and $m_b=5.10\pm0.40$ GeV. The cross sections for $E_{cm}=m_Z$ with varying $m_c$ and $m_b$ are presented in Tables~\ref{uncertaintyz0c} and \ref{uncertaintyz0b}, which are more sensitive to $m_c$ than those of $m_b$. Here we only consider the cross sections for the channels through $Z^{0}$-propagator, since the channels through $\gamma$-propagator is small and negligible as shown by Table~\ref{crosssection}. By adding these two uncertainties in quadrature, we obtain
\begin{eqnarray}
\sigma_{e^++e^-\to Z^{0} \to \Xi_{cc}([^3S_1]_{\bf\bar{3}}) +\bar{c}\bar{c}} &=& 0.727^{-0.275}_{+0.543} \;{\rm pb},\\
\sigma_{e^++e^-\to Z^{0} \to \Xi_{cc}([^1S_0]_{\bf 6}) +\bar{c}\bar{c}} &=& 0.353^{-0.133}_{+0.263} \;{\rm pb},\\
\sigma_{e^++e^-\to Z^{0} \to \Xi_{bc}([^3S_1]_{\bf\bar{3}}) +\bar{b}\bar{c}} &=& 1.00^{-0.402}_{+0.850} \;{\rm pb},\\
\sigma_{e^++e^-\to Z^{0} \to \Xi_{bc}([^3S_1]_{\bf 6}) +\bar{b}\bar{c}} &=& 0.502^{-0.203}_{+0.424} \;{\rm pb},\\
\sigma_{e^++e^-\to Z^{0} \to \Xi_{bc}([^1S_0]_{\bf\bar{3}}) +\bar{b}\bar{c}} &=& 0.730^{-0.275}_{+0.55} \;{\rm pb},\\
\sigma_{e^++e^-\to Z^{0} \to \Xi_{bc}([^1S_0]_{\bf 6}) +\bar{b}\bar{c}} &=& 0.365^{-0.138}_{+0.274} \;{\rm pb},\\
\sigma_{e^++e^-\to Z^{0} \to \Xi_{bb}([^3S_1]_{\bf\bar{3}}) +\bar{b}\bar{b}} &=& 0.080^{-0.018}_{+0.025} \;{\rm pb},\\
\sigma_{e^++e^-\to Z^{0} \to \Xi_{bb}([^1S_0]_{\bf 6}) +\bar{b}\bar{b}} &=& 0.039^{-0.009}_{+0.012} \;{\rm pb}.\\ \nonumber
\end{eqnarray}

\section{summary}

Using the NRQCD factorization formula, we have studied the production of the doubly heavy baryon $\Xi_{QQ^{\prime}}$ ($\Xi_{cc}$, $\Xi_{bc}$, $\Xi_{bb}$) through the $e^+e^-$ annihilation. When the $e^+ e^-$ collision energy is around the $Z^0$-peak, sizable cross-sections for the doubly heavy baryon can be obtained. Contributions from those $(QQ^{\prime})[n]$-diquark states with the same importance have been discussed. By adding all these intermediate diquark states' contributions together, we obtain the total cross section for $\Xi_{QQ^{\prime}}$ at $E_{cm}=Z^{0}$ :
\begin{eqnarray}
\sigma_{e^+ e^-\to \gamma/Z^{0} \to \Xi_{cc} +\bar{c}\bar{c}} &=& 1.08^{-0.409}_{+0.806} \;{\rm pb},\\
\sigma_{e^+ e^-\to \gamma/Z^{0} \to \Xi_{bc} +\bar{b}\bar{c}} &=& 2.60^{-1.02}_{+2.10} \;{\rm pb},\\
\sigma_{e^+ e^-\to \gamma/Z^{0} \to \Xi_{bb} +\bar{b}\bar{b}} &=& 0.119^{-0.027}_{+0.037} \;{\rm pb},
\end{eqnarray}
where the errors are caused by varying $m_c=1.80\pm0.30$ GeV and $m_b=5.10\pm0.40$ GeV.

At the super $Z$-factory with a high luminosity up to ${\cal L}\propto 10^{34} - 10^{36} cm^{-2}s^{-1}$, one would expect to accumulate $1.1\times10^{5 \sim 7}$ $\Xi_{cc}$ events,  $2.6\times10^{5 \sim 7}$ $\Xi_{bc}$ events and  $1.2\times 10^{4 \sim 6}$ $\Xi_{bb}$ events in one operation year. If taking the $e^+ e^-$ collision energy to run slightly off the $Z^{0}$-peak, i.e. $\sqrt{S}=0.95 m_Z$ or $1.05 m_Z$, the total production cross section will be lowered by about one order of magnitude from its peak value. However, one may still observe sizable events. At the hadronic colliders, there is much pollution from the hadronic background and many produced baryon events shall be cut off by the trigging condition. So, some alternative measurements would be helpful for a comprehensive study. In addition to the hadronic collider LHC, the super Z-factory will provide another good platform for studying the $\Xi_{QQ^{\prime}}$ baryon properties.

As a final remark, if not too much baryon events are cut off by the trigging condition at the super Z-factory, one may have the chance to further distinguish the baryons with different light quark content. When an intermediate diquark is formed, it will grab a light quark (with gluons if necessary) from the `environment' to form a colorless doubly heavy baryon with a relative possibility for various light quarks as $u :d :s \simeq 1:1:0.3$~\cite{pythia}. For example, if the diquark $(cc)[^3S_1]_{\bar{3}}$ is produced, then it will fragment into $\Xi_{cc}^{++}$ with $43\%$ probability, $\Xi_{cc}^{+}$ with $43\%$ probability and $\Omega_{cc}^{+}$ with $14\%$ probability accordingly. The condition is the same for the production of $\Xi^{0}_{bb}$, $\Xi^{-}_{bb}$ and $\Omega^{-}_{bb}$ or the production of $\Xi^{+}_{bc}$, $\Xi^{0}_{bc}$ and $\Omega^{0}_{bc}$.

\hspace{2cm}

{\bf Acknowledgements}: This work was supported in part by the Fundamental Research Funds for the Central Universities under Grant No.CDJXS11100011, the Program for New Century Excellent Talents in University under Grant No.NCET-10-0882, and the Natural Science Foundation of China under Grant No.11075225.

\end{document}